\documentclass[runningheads]{llncs}
\usepackage{graphicx}
\usepackage{tikz}
\usepackage{tkz-graph}
\usepackage{bbm}
\usepackage{booktabs}
\usepackage{listings}
\usepackage{amsfonts}
\usepackage{amsmath}
\usepackage{wrapfig}

\usepackage{subfig}

\newtheorem{mydef}{Definition}

\begin{document}
\title{StarStar Models: Process Analysis on top of Databases}
%
%
\author{Alessandro Berti\inst{1}\orcidID{0000-0003-1830-4013} \and
Wil van der Aalst\inst{1}\orcidID{0000-0002-0955-6940}}
%
%
\institute{Process and Data Science department, Lehrstuhl fur Informatik 9 52074 Aachen, RWTH Aachen University, Germany}
\maketitle              
\begin{abstract}
Much time in process mining projects is spent on finding and understanding data sources and extracting the event data needed. As a result, only a fraction of time is spent actually applying techniques to discover, control and predict the business process.
Moreover, there is a lack of techniques to display relationships on top of databases without the need to express a complex query to get the required information.
In this paper, a novel modeling technique that works on top of databases is presented. This technique is able to show a multigraph representing activities inferred from database events, connected with edges that are annotated with frequency and performance information.
The representation may be the entry point to apply advanced process mining techniques that work on classic event logs, as the model provides a simple way to retrieve a classic event log from a specified piece of model.
Comparison with similar techniques and an empirical evaluation are provided.

\keywords{Process Mining \and Database Querying.}
\end{abstract}
\section{Introduction}
\label{sec:introduction}
\begin{figure}[ht]
\centering
\subfloat[Classical ETL scenario]{{ \includegraphics[height=185px]{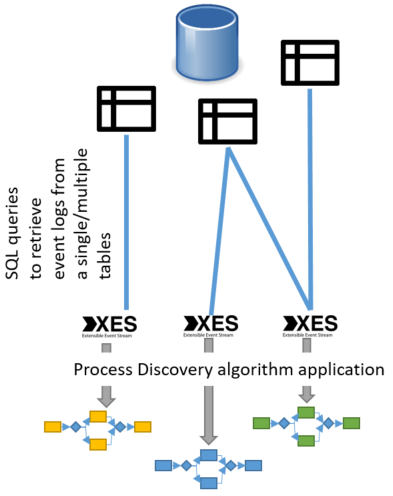} }}
\qquad
\subfloat[StarStar models]{{ \includegraphics[height=185px]{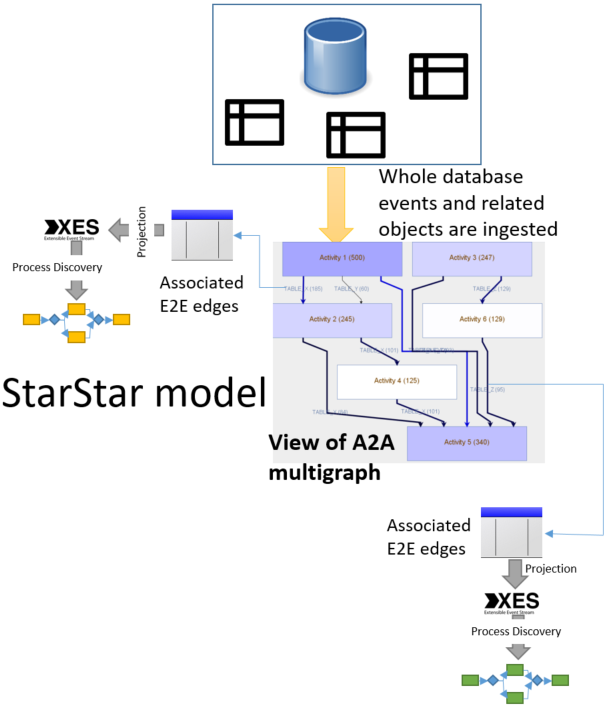} }}
\caption{Comparison between two different process mining ETL scenarios: the first is about choosing some views on the data and getting
several event logs from the same database that could be used to obtain different process models using mainstream discovery techniques.
The second is about inserting all the data in a StarStar model and observing the processes on the A2A multigraph having also the possibility (drill down) to choose a case notion.
When a case notion is selected, a classic event log can be generated and subsequently analyzed by existing process mining techniques.
}
\label{fig:approachIntro}
\end{figure}
Process mining is a growing branch of data science that aims to extract insights from event data recorded in information systems. Several techniques exist to analyze event data:
process discovery algorithms are able to find a process model, conformance checking compare event data with a process model to find deviations, predictive algorithms aim to predict
the future of an incomplete case based on historical data. Gathering event data is, therefore, a very important aspect for the success of process mining projects, that start
from the retrieval and the validation of an event log, i.e., a log of events related to different executions of a business process.
Information systems are often built on top of relational databases, to ensure data integrity and normalization. Relational databases contain entities (tables) and relations
between entities; events could be recorded in tables of the databases (through {\it in-table versioning}) or could be retrieved using some database log (like {\it redo logs}).
In both cases, to extract an event log a {\it case notion} should be chosen: this means that a specific view on the database is selected, events related to this view are retrieved
from the database and grouped by a specific set of columns (the {\it case id}) into cases. Several views on the database could be retrieved, this means that for the same database
several event logs and process models could be extracted.
However, if the process is very complex, involving many different entities, the construction of the view is not easy: 
deep knowledge of the process and the table structure may be required, moreover the execution time of queries needed to extract event data may be very time consuming. This approach of querying
event data is represented on the left-hand side of Fig. \ref{fig:approachIntro}.
There has been some research regarding the possibility to make SQL queries easier \cite{calvanese2017ontop,bouchou2014semantic}. The basic idea is to provide the business analyst a way to express queries in a simpler language (SPARQL query).
Recently two other approaches have been proposed in order to support process mining on databases:
\begin{itemize}
\item OpenSLEX meta-models \cite{de2018connecting}. In this case, the database updates are inserted into a meta-model where events, activities associated to events, objects, object versions and relationships are stored in a clear way
and can be retrieved with simple queries. To obtain classic event logs, a case notion (connecting events each other) needs to be used.
\item Object-centric models \cite{li2017automatic2}. In this case, a process model involving entities and relations in the ER model, connecting activities related to events,
is found, and multiple case notions may coexist. This class of models provides scope for conformance checking application and can be used to learn interesting patterns related to the cardinality of events.
\end{itemize}
\begin{figure}
\begin{minipage}[c]{0.5\textwidth}
\includegraphics[height=250px]{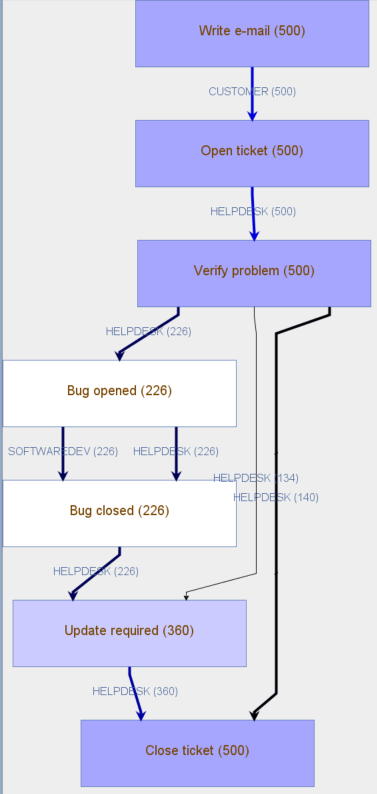}
\end{minipage}\hfill
\begin{minipage}[c]{0.5\textwidth}
\caption{Representation of the activities multigraph in a StarStar model of a software house ticketing system. The management of customer problems involves events of several different entities (e.g., ticket management, delivery appointment registration, software bug lifecycle management)
but could be captured in a unified way by the activities multigraph.}
\label{fig:softwareHouse}
\end{minipage}
\end{figure}
There are some difficulties to successfully apply these approaches in real-life applications.
OpenSLEX provides indeed some connectors for database logs (redo logs, in-table versioning, or specific database formats \cite{ingvaldsen2007preprocessing}), but require the specification of a single case notion in order to get an event log.
Although OpenSLEX meta-models are simpler than the actual database,
it is difficult to select the right case notion that may depend on the question to answer. Once the case notion is expressed, it is easy to get from the meta-model a proper event log to apply mainstream process mining techniques.
Object-centric models, after the discovery algorithm has been applied, provide the user a visualization of a model that involves entities and activities, and this does not require the specification
of a case perspective, but a classic event log cannot be obtained (and so existing process mining techniques could not be applied).
This paper introduces a new modeling technique that is able to calculate a graph where relationships between activities are shown without forcing the user to specify a case notion,
since different case notions are combined in one succint diagram.

The resulting models are called StarStar models. Such models belong to the class of artifact-centric models \cite{cohn2009business,nigam2003business} that combine
data and process in a holistic manner to discover patterns and check compliance \cite{lohmann2011compliance}.
StarStar models provide a graphical way to choose a case notion and analyze an event log to use with classic process mining techniques.

A {\it StarStar model} is a representation of event data contained in a database. It is a collection of several graphs:
\begin{itemize}
\item An {\it event to object graph} $\textrm{E2O}$ that aims to represent events and objects extracted from the database and relationships between them.
\item An {\it event to event multigraph} $\textrm{E2E}$ that aims to represent directly-follows relationships between events in the perspective of some object.
\item An {\it activities multigraph} $\textrm{A2A}$ that aims to represent directly-follows relationships between activities in the perspective of some object class.
\end{itemize}
ETL scenario using StarStar models is shown on the right-hand side of Fig. \ref{fig:approachIntro}.
The visualization part of a StarStar model is able to show a multigraph between activities (A2A); however, relations in the E2O and E2E multigraphs are important for filtering the model and for performing a projection
on the selected case notion. The E2O graph is obtained directly from the data. For the E2E and the A2A multigraphs some algorithms (that runs in linear complexity) will be provided in the following sections.
An example of a StarStar model representing a software house ticketing system is shown in Fig. \ref{fig:softwareHouse}.
The approach has been implemented as a plugin in the ProM tool and evaluated using data sets extracted from a Customer Relationship Management (CRM) system.

\section{Background}
\label{sec:background}
Relational databases are organized in entities (classes of {\it objects} sharing some properties), relationships (connections between entities \cite{chen1976entity}), attributes (properties of entities and relationships).
Events can be viewed as updates of a database (e.g. insertion of new objects, changes to existing objects, removal of existing objects).
Some ways to retrieve events from databases are:
\begin{itemize}
\item Using redo logs (see \cite{de2018connecting}). These are logs where each operation in the database is saved with a timestamp; this helps to guarantee consistency, and possibility to
rollback and recovery.
\item Using in-table versioning. In this case, the primary key is enriched by a timestamp column. For each phase of the lifecycle of an object, a new entry is added to the in-table
versioning, sharing the primary key values except the timestamp column.
\end{itemize}
An event may be linked to several objects (for example, the event that starts a marketing campaign in a CRM system may be linked to several customers), and an object may be linked to several events
(for example, each customer can be related to all the tickets it opens). For the following definition, let $\mathcal{U}_E$ be the universe of events (all the events happening in a database context),
$\mathcal{U}_O$ the universe of objects (all the objects that are instantiated in the database context),
$\mathcal{U}_{OC}$ is the universe of object classes (a class define the structure and the behavior of a set of objects),
$\mathcal{U}_A$ the universe of activities (names referring to a particular step of a process),
$\mathcal{U}_{\textrm{attr}}$ the universe of attribute names (all the names of the attributes that can be related to an event),
$\mathcal{U}_{\textrm{val}}$ the universe of attribute values (all the possible values for attributes).
It is possible to define a function $\textrm{class} : \mathcal{U}_O \rightarrow \mathcal{U}_{OC}$ that associates each object to the corresponding object class.
\begin{mydef}[Event log in a database context]
\label{def:newEventLogDefinition}
An event log in a database context is a tuple $L_D = (E, \textrm{act}, \textrm{attr}, \allowbreak \textrm{EO}, \leq)$ where:
\begin{itemize}
\item $E \subseteq \mathcal{U}_E$ is a set of events
\item $\textrm{act} \in E \rightarrow \mathcal{U}_A$ maps events onto activities
\item $\textrm{attr} \in E \rightarrow (\mathcal{U}_{attr} \not\rightarrow \mathcal{U}_{val})$ maps events onto a partial function assigning values to some attributes
\item $\textrm{EO} \subseteq E \times \mathcal{U}_O$ relates events to sets of object references
\item $\leq ~ \subseteq E \times E$ defines a total order on events.
\end{itemize}
\end{mydef}
An example attribute of an event $e$ is the timestamp $\textrm{attr}(e)(time)$ which refer to the time the event happened.
Given the previous definition, we could get the set of objects referenced in the event log as $O = \{ o ~ | ~ (e, o) \in \textrm{EO} \}$.
For the classic process mining event log definition, let $\mathcal{U}_C$ be the universe of case identifiers.
\begin{mydef}[Classic event log]
\label{def:classicalEventLogDefinition}
An event log is a tuple $L = (C, E, \textrm{case\_ev}, \textrm{act}, \allowbreak \textrm{attr}, \leq)$ where:
\begin{itemize}
\item $C \subseteq \mathcal{U}_C$ is a set of case identifiers
\item $E \subseteq \mathcal{U}_E$ is a set of events
\item $\textrm{case\_ev} \in C \rightarrow \mathcal{P}(E)$ maps case identifiers onto set of events (belonging to the case)
\item $\textrm{act} \in E \rightarrow \mathcal{U}_A$ maps events onto activities
\item $\textrm{attr} \in E \rightarrow (\mathcal{U}_{attr} \not\rightarrow \mathcal{U}_{val})$ maps events onto a partial function assigning values to some attributes
\item $\leq ~ \subseteq E \times E$ defines a total order on events.
\end{itemize}
\end{mydef}
To project an event log in a database context to a classic event log, a case notion needs to be chosen, so events
that should belong to the same case can be grouped.
Let $L_D = (E, \textrm{act}, \textrm{attr}, \textrm{EO}, \leq)$ be an event log in a database context. A {\it case notion} is a set $C_D \subseteq \mathcal{P}(E) \setminus \emptyset$ such that $\bigcup_{x \in C_D} x = E$.
Let $\textrm{id} : C_D \rightarrow \mathcal{U}_C$ a function that associates a case identifier to a set of events.
It is possible to define a projection function:
$$\textrm{proj}(C_D,L_D) = (C,E,\textrm{case\_ev},\textrm{act},\textrm{attr},\leq)$$
where $C = \cup_{x \in C_D} \textrm{id}(x)$, $\textrm{case\_ev} \in C \rightarrow \mathcal{P}(E)$ such that for all $c \in C_D$, $\textrm{case\_ev}(\textrm{id}(c)) = c$.
An implementation of the event log format described in Definition \ref{def:newEventLogDefinition} is the XOC format \cite{li2017automatic2},
that is expressed in XML language.
\section{Approach}
In this section, a definition of the components of StarStar models will be introduced. The E2O graph will be obtained directly from the database logs; the E2E multigraph will be
obtained in linear complexity by calculating directly-follows relationships between events in the perspective of some object; the A2A multigraph will be obtained in linear complexity
by calculating directly-follows relationships between activities in the perspective of some object class, using the information stored in the E2E multigraph. The A2A multigraph
is the visual element of a StarStar model, and a projection function will be introduced in this section to obtain a classic event log when a perspective is chosen.

\subsection{Construction of the Model}
\label{sec:modelConstruction}
StarStar models are composed of several graphs (E2O, E2E, A2A) and are constructed by reading a representation of event data retrieved from a database (OpenSLEX meta-models or XOC logs).
\begin{mydef}[E2O graph]
Let $L_D = (E, \textrm{act}, \textrm{attr}, \textrm{EO}, \leq)$ be an event log in a database context.
$\textrm{EO} \subseteq E \times O$ is an event to object graph relating events ($E$) and objects ($O$). 
\label{def:e2ograph}
\end{mydef}
The E2O graph is obtained directly from the data without any transformation. The remaining steps in the construction
of a StarStar model are the construction of the E2E multigraph and of the A2A multigraph.
The following functions on objects are used:
\begin{itemize}
\item $g : \mathcal{U}_O \rightarrow \mathcal{P}(\mathcal{U}_E)$, $g(o) = \{ e \in \mathcal{U}_E ~ \arrowvert ~ (e, o) \in \textrm{EO} \}$ is a function that for each object returns the set of events that are
related to the object.
\item $w : \mathcal{U}_O \rightarrow \mathbb{R}$, $w(o) = \frac{1}{\arrowvert g(o) \arrowvert + 1}$ is the weight of the object and is defined as the inverse of the cardinality of the set of related events to the given object plus $1$. This weight function will be used in the construction of E2E multigraph to give less importance to relations based on objects that are related to many events.
\item $\sharp_k : \mathcal{U}_O \rightarrow \mathcal{U}_E$, $\sharp_k(o) = e \quad \textrm{such that} \quad e \in g(o) ~ \wedge ~ \arrowvert \{ e' \in g(o) ~ | ~ e' \leq e \} \arrowvert = k$ for $1 \leq k \leq \arrowvert g(o) \arrowvert$ is a function that in the totally ordered set $g(o)$ returns the $k$-th element.
\end{itemize}
\begin{mydef}[E2E multigraph]
Let $L_D = (E, \textrm{act}, \textrm{attr}, \textrm{EO}, \leq)$ be an event log in a database context. Let
$$F_E = \{ (o, i) ~ \arrowvert ~ o \in O ~ \wedge ~ 2 \leq i \leq \arrowvert g(o) \arrowvert \}$$
such that for every edge $f_E \in F_E$ the following attributes are defined:
\begin{itemize}
\item $\Pi^{E}_{\textrm{obj}}(f_E) \in O$ is the object associated to the edge.
\item $\Pi^{E}_{\textrm{in}}(f_E) \in E$ is the input event associated to the edge.
\item $\Pi^{E}_{\textrm{out}}(f_E) \in E$ is the output event associated to the edge.
\item $\Pi^{E}_{\textrm{weight}}(f_E) \in \mathbb{R}^{+}$ associates each edge to a positive real number expressing its weight.
\item $\Pi^{E}_{\textrm{perf}}(f_E) \in \mathbb{R}^{+} \cup \{ 0 \}$ associates each edge to a non-negative real number expressing its performance.
\end{itemize}
For $f_E = (o, i) \in F_E$:
\begin{itemize}
\item $\Pi^{E}_{\textrm{obj}}(f_E) = o$
\item $\Pi^{E}_{\textrm{in}}(f_E) = \sharp_{i-1}(o)$
\item $\Pi^{E}_{\textrm{out}}(f_E) = \sharp_i(o)$
\item $\Pi^{E}_{\textrm{weight}}(f_E) = w(o)$
\item $\Pi^{E}_{\textrm{perf}}(f_E) = \textrm{attr}(\Pi_{\textrm{out}}(f_E))(time) - \textrm{attr}(\Pi_{\textrm{in}}(f_E))(time)$
\end{itemize}
The event to event multigraph (E2E) can be introduced having events as nodes and associating each couple of events $(e_1, e_2) \in E \times E$ to the following set of edges:
$$R_E(e_1, e_2) = \left \{ f_E \in F_E ~ \arrowvert ~ \Pi^{E}_{\textrm{in}}(f_E) = e_1 \wedge \Pi^{E}_{\textrm{out}}(f_E) = e_2 \right \}$$
\end{mydef}
The introduction of the set $F_E$ is useful for the definition of the A2A multigraph (that is the visual element of a StarStar model). Although it does not make sense to represent the overall E2E multigraph, involving relationships between all events, it may be useful
to display directly-follows relationships involving a specific event or a specific couple of events. In this case, a representation of the E2E multigraph draws as many edges between a couple of events $(e_1, e_2) \in E \times E$ as the number
of elements contained in the set $R_E(e_1, e_2)$. To each edge $f_E \in R_E(e_1, e_2)$, a label could be associated in the representation taking as example the weight $\Pi^{E}_{\textrm{weight}}(f_E)$
or the performance $\Pi^{E}_{\textrm{perf}}(f_E)$.

\begin{mydef}[A2A multigraph]
Let $L_D = (E, \textrm{act}, \textrm{attr}, \textrm{EO}, \leq)$ be an event log in a database context. Let
$$F_A = \{ (c, (a_1, a_2)) ~ \arrowvert ~ c \in \mathcal{U}_{OC} ~ \wedge ~ (a_1, a_2) \in \mathcal{U}_A \times \mathcal{U}_A \}$$
such that for each edge $f_A \in F_A$ the following attributes are defined:
\begin{itemize}
\item $\Pi^{A}_{\textrm{class}}(f_A) \in \mathcal{U}_{OC}$ is the class associated to the edge.
\item $\Pi^{A}_{\textrm{in}}(f_A) \in \mathcal{U}_A$ is the source activity associated to the edge.
\item $\Pi^{A}_{\textrm{out}}(f_A) \in \mathcal{U}_A$ is the target activity associated to the edge.
\item $\Pi^{A}_{\textrm{count}}(f_A) \in \mathbb{N}$ associates each edge to a natural number expressing the number of occurrences.
\item $\Pi^{A}_{\textrm{weight}}(f_A) \in \mathbb{R}^{+}$ associates each edge to a positive real number expressing its weight.
\item $\Pi^{A}_{\textrm{perf}}(f_A) \in \mathbb{R}^{+} \cup \{ 0 \}$ associates each edge to a non-negative real number expressing its performance.
\end{itemize}
Let $\textrm{AE} : F_A \rightarrow \mathcal{P}(F_E)$ be a function such that for $f_a \in F_A$: \\
$\textrm{AE}(f_A) = \{ ~ f_E \in F_E ~ \arrowvert ~ \textrm{class}(\Pi^{E}_{\textrm{obj}}(f_E)) = \Pi^{A}_{\textrm{class}}(f_A) ~ \wedge \allowbreak ~ \textrm{act}(\Pi^{E}_{\textrm{in}}(f_E)) = \Pi^{A}_{\textrm{in}}(f_A) ~ \wedge \allowbreak ~ \textrm{act}(\Pi^{E}_{\textrm{out}}(f_E)) = \Pi^{A}_{\textrm{out}}(f_A) ~ \}$. Then for $f_A = (c, (a_1, a_2)) \in F_A$:
\begin{itemize}
\item $\Pi^{A}_{\textrm{class}}(f_A) = c$
\item $\Pi^{A}_{\textrm{in}}(f_A) = a_1$
\item $\Pi^{A}_{\textrm{out}}(f_A) = a_2$
\item $\Pi^{A}_{\textrm{count}}(f_A) = \arrowvert \textrm{AE}(f_A) \arrowvert$
\item $\Pi^{A}_{\textrm{weight}}(f_A) = \sum_{f_E \in \textrm{AE}(f_A)} \Pi^{E}_{\textrm{weight}}(f_E)$
\item $\Pi^{A}_{\textrm{perf}}(f_A) = \frac{\sum_{f_E \in \textrm{AE}(f_A)} \Pi^{E}_{\textrm{perf}}(f_E)}{\Pi^{A}_{\textrm{count}}(f_A)}$
\end{itemize}
The activities multigraph (A2A) can be introduced having activities as nodes and associating each couple of activities $(a_1, a_2) \in A \times A$ to the following set of edges:
$$R_A(a_1, a_2) = \left \{ f_A \in F_A ~ \arrowvert ~ \Pi^{A}_{\textrm{in}}(f_A) = a_1 \wedge \Pi^{A}_{\textrm{out}}(f_A) = a_2 \right \}$$
\end{mydef}
A representation of the A2A multigraph (that is the visual element of a StarStar model) draws as many edges between a couple of activities $(a_1, a_2) \in A \times A$ as the number of elements contained in the set $R_A(a_1, a_2)$.
To each edge $f_A \in R_A(a_1, a_2)$, a label could be associated in the representation taking as example the number of occurrences $\Pi^{A}_{\textrm{count}}(f_A)$, the weight $\Pi^{A}_{\textrm{weight}}(f_A)$
or the performance $\Pi^{A}_{\textrm{perf}}(f_A)$.
Since by construction the edges in this graph can be associated to elements in the E2E graph (through the $\textrm{AE}$ function), the possibility to drill down to a classic event log (choosing a case notion) is maintained.



\subsection{Projection to a classic event log}
\label{sec:projectionExpl}
In Section \ref{sec:background}, the log concept used in this paper, that is different from the classic one, has been proposed and a projection function that transforms events from the first format to the second has been introduced.
A2A graphs have by construction the functionality to drill down a specific class perspective and keep events related to this perspective, so it is provided the possibility to get a classic event log
out of a StarStar model.
Let $L_D = (E, \textrm{act}, \textrm{attr}, \textrm{EO}, \leq)$ be an event log in a database context.
The following similarity function between sets of events could be introduced:
$$\textrm{sim} : \mathcal{P}(\mathcal{U}_E) \times \mathcal{P}(\mathcal{U}_E) \rightarrow \mathbb{R} \qquad \textrm{sim}(E_1, E_2) = \frac{\arrowvert E_1 \cap E_2 \arrowvert}{\textrm{max}(\arrowvert E_1 \arrowvert, \arrowvert E_2 \arrowvert)}$$
along with the following case notion (a class $c \in \mathcal{U}_{OC}$ is chosen):
$$C^0 : \mathcal{U}_{OC} \times (0, 1] \rightarrow \mathcal{P}(E)$$
$$C^0(c, \omega) = \bigcup_{o_1 \in O, \textrm{class}(o_1) = c} \bigcup_{~ ~ o_2 \in O, \textrm{sim}(g(o_1), g(o_2)) \ge \omega} g(o_2) \cup g(o_1)$$
Then a classical event log could be obtained as $\textrm{proj}(L_D, C^0(c, \omega))$. When $\omega = 1$, then the case notion takes
the set of related events for each object belonging to the class. When $\omega$ is near to $0$, for example $\omega = 0.01$, then cases are formed by the union of several sets of events. The parameter $\omega$ is
called {\it connection weight threshold}. We could define iteratively different case notions (for $i \ge 1$):
$$C^i : \mathcal{U}_{OC} \times (0, 1] \rightarrow \mathcal{P}(E)$$
$$C^i(c, \omega) = \bigcup_{o_1 \in O, \textrm{class}(o_1) = c} \bigcup_{~ ~ s \in C^{i-1}(c, \omega), \textrm{sim}(g(o_1), s) \ge \omega} s \cup g(o_1)$$
Then a classical event log could be obtained as $\textrm{proj}(L_D, C^i(c, \omega))$. The number of iterations $i$ is called {\it log window}. With a connection weight threshold
that is relatively near to $0$, the size of each case increases at the increase of the number of iterations done to define the case notion.

\section{Support tool}
\label{sec:support}
In order to evaluate the StarStar models representation, a ProM plug-in has been realized that is able to take as input a representation of the events happening at database
level, is able to calculate the StarStar model starting from the data and to show it to the end user using the mxGraph library. The supported input data types include:
\begin{itemize}
\item XOC logs \cite{li2017automatic2}, that have been presented in Section \ref{sec:background}, are XMLs storing events along with their related objects and the status of the object model
at the time the event happened. These can be imported from the Import button by providing the XOC log. Some XOC logs
could be found in the {\it tests} folder of the project. In particular, the following logs are provided:
\begin{itemize}
\item An example ERP log ({\it erp.xoc}) extracted from a Dollibar CRM installation. The resulting multigraph is very simple and consists of the following perspectives:
{\it supplier\_order\_line}, {\it dispatch}.
\item A commercial opportunities log ({\it opportunities.xoc}) extracted from a Dynamics CRM demo installation. The multigraph contains many perspectives and activities.
\end{itemize}
\begin{figure}[ht]
\centering
\includegraphics[width=325px]{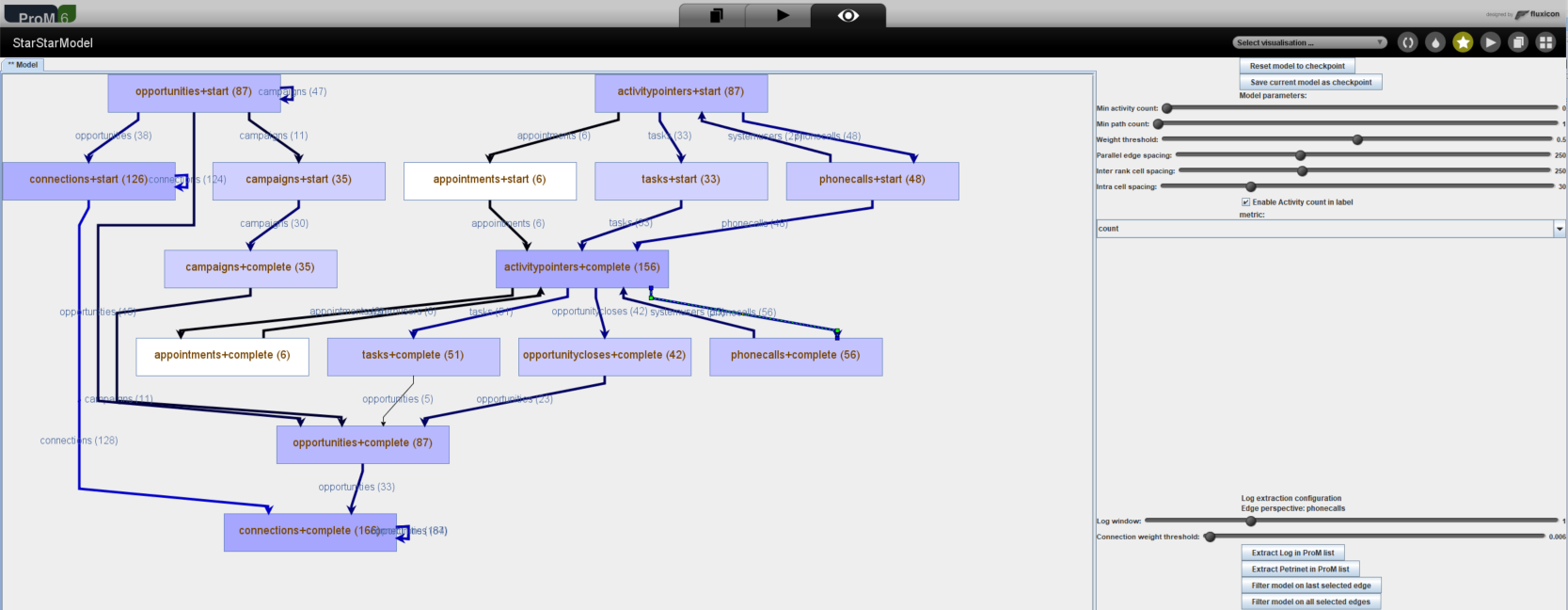}
\caption{StarStar models visualization plug-in inside the ProM framework.}
\label{fig:overallPlugin}
\end{figure}
After reading the meta-model, the E2O graph is built in-memory, and so the E2E graph and the A2A graph. After importing, the model can be viewed by clicking the View button in ProM.
\item OpenSLEX meta-models \cite{de2018connecting}. As example, we provide an OpenSLEX \allowbreak meta-model \allowbreak {\it metamodel.slexmm} generated from a \allowbreak concert ticket management \allowbreak database. This can be imported, in a similar way to XOC logs, using the Import button in the ProM interface.
As for XOC logs, the E2O graph is built in-memory, and so the E2E graph and the A2A graph, and the model can be viewed by clicking the View button in ProM.
\item Neo4J databases. In this case, event and object data extracted from a database are inserted in a prior step as nodes in the graph, and edges are inserted in the
graph if there is a relationship between an event and an object. This is the E2O graph. The E2E graph is formed by calculating the directly-follows relation on the E2O graph
(this can be done executing a script on the database), and the A2A graph is formed starting from the E2E graph by aggregating connections as explained in Section \ref{sec:modelConstruction}.
Then, the ProM framework can provide visualization of the StarStar models by querying Neo4J and doing specific queries. The plug-in {\it StarStarModel Neo4J Connector} implements the
connector to this data source by simply providing the hostname and the port of the database.
\end{itemize}
\begin{figure}[ht]
\centering
\includegraphics[width=250px]{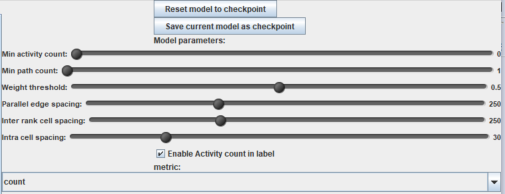}
\caption{Representation of the sliders provided in the ProM plug-in: {\it Minimum activity count} regulates the minimum number of occurrences of an activity displayed in the multigraph;
{\it Minimum path count} regulates the minimum number of occurrences of a path displayed in the multigraph;
{\it Weight threshold} regulates the threshold to use to keep/filter out edges in the displayed graph;
{\it Parallel edges spacing} specifies the distance between parallel edges in the same view;
{\it Inter rank cell spacing} specifies the distance between ranks (different hierarchies of nodes as discovered by the layouting algorithm);
{\it Intra cell spacing} specifies the distance between cells of the same rank (same hierarchy in the layouting algorithm).
Another feature that has been implemented in order to improve usability are the checkpoints: during a filtering process, it may be useful to save a status and retrieve it later. The two buttons
{\it Reset model to checkpoint} and {\it Save current model as checkpoint} are provided on the top of the right panel.
}
\label{fig:toolSliders}
\end{figure}
\begin{figure}[ht]
\centering
\includegraphics[width=250px]{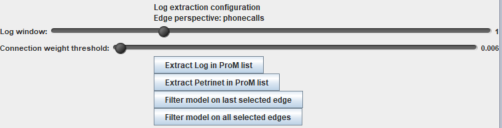}
\caption{Log projection functionality implemented in the ProM plug-in. The parameters are the log window, that is the number of iterations done by the projection algorithm, and the connection weight threshold, that is the minimum connection strength between two objects that should exist to enable the merge between the sets of related events.
The projection plug-in provides the possibility to extract the event log that is stored in the ProM workspace, or to apply directly the Inductive Miner to
discover a process schema from the projected log and open it as a tab of the visualization.}
\label{fig:projection}
\end{figure}
The ProM implementation of the importer for XOC logs and OpenSLEX meta models works in-memory. However, the Neo4J database scales better, and is possible to upload an XOC log
into a Neo4J database and, after that, querying it. This part has been implemented in Python and is available in the {\it xocNeo4JUpload} folder.
The plug-in (shown in Fig. \ref{fig:overallPlugin}) provides a tabbed view of the A2A multigraph on the left, along with some sliders on the right regulating the complexity of the graph displayed.
Since the A2A multigraph may contain many edges, an initial set-up of the visualization is made in order to keep only edges between activities that have a meaningful weight.
As a result of removing such edges, activities may become disconnected.
With the default selection of the metric, activities are decorated by a color that is dark when the number of occurrences of the activity is high, and their label is the
name of the activity along with the number of occurrences of the activity. Edges color and width depend on the frequency of the relation, and the label reports the perspective
and the number of occurrences of the relation. Some sliders, shown in Fig. \ref{fig:toolSliders}, are provided to the user in order to regulate the graph visualization.
Since several metrics could be projected into the arcs, a metric selection menu has been implemented on the right panel in order to choose the metric (count, performance, weight).
The color of activities and arcs, and the width of the arcs, depend on the metric and on the value of the arc. For example, when the frequency metric is selected, then the arc has greater width when
the number of occurrences of the relation described by the edge is higher.
The visualization plug-in provides several options to filter the StarStar model and keep only activities/perspectives that are interesting.
In particular, selecting some edges by using the CTRL+Click control, it is possible to apply a filter on edges. The filter works as follows: all objects belonging to the perspective and related
to events which activity is the source activity of the edge are retrieved into an object set, and then the StarStar model is calculated again keeping only events that are related
to at least one object in the object set. This is similar, in the classic process discovery setting, to filtering in a log all the events belonging to cases where an edge between activities
is present at least once, and is similar to the approaches described in \cite{catarci1997visual,chau2008graphite}.
Then, for edges is provided the possibility to retrieve a classic event log, in the way explained in Section \ref{sec:projectionExpl} and shown in Fig. \ref{fig:projection}.

\section{Evaluation}
\label{sec:evaluation}
\begin{figure}[ht]
\centering
\includegraphics[width=338px]{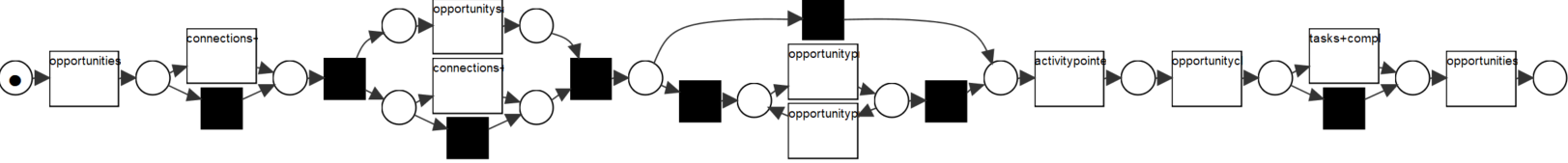}
\caption{Representation of the Petri net obtained choosing the {\it opportunity} perspective on the graph and applying projection.}
\label{fig:petriNet}
\end{figure}

\begin{figure}[ht]
\centering
\includegraphics[width=338px]{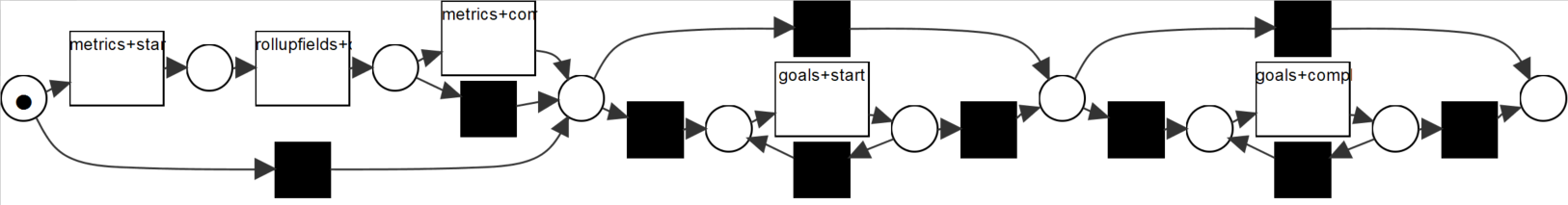}
\caption{Representation of the Petri net obtained choosing the {\it metrics} perspective on the graph and applying projection.}
\label{fig:petriNet2}
\end{figure}

This section presents a study of data extracted from a Microsoft Dynamics CRM demo and analyzed through a StarStar model. A Customer Relationship Management system (CRM) \cite{buttle2004customer} is an information
system used to manage the commercial lifecycle of an organization, including management of customers, opportunities and marketing campaigns.
Many companies actually involve a CRM system for helping business and sales people to coordinate, share information and goals.
Data extracted from Microsoft Dynamics CRM is particularly interesting since this product manages several processes of the business side, providing the possibility to define workflows
and to measure KPI also through connection to the Microsoft Power BI business intelligence tool.
For evaluation purposes, an XOC log has been generated containing data extracted from a Dynamics CRM demo.
The database supporting the system contains several entities, and each entity contains several entries related to activities happening in the CRM.
Each entry could be described by a unique identifier (UUID), the timestamp of creation/change, the resource that created/modified the entry, and some UUIDs of other
entries belonging to the same or to different entities. Moreover, each entry is uniquely associated to the entity it belongs to.
The following strategy has been pursued in order to generate an XOC log:
\begin{itemize}
\item For each entry belonging to an entity, two events have been associated: creation event (with the timestamp of creation and lifecycle {\it start}) and modify event
(with the timestamp of modification and lifecycle {\it complete}).
\item Each entry belonging to an entity has also been associated to an object.
\item Relationships between events and objects are created accordingly to the relationships expressed by the entries (an entry may cite several UUIDs of other entries stored in the database).
\end{itemize}
\begin{figure}[ht]
\centering
\includegraphics[width=345px]{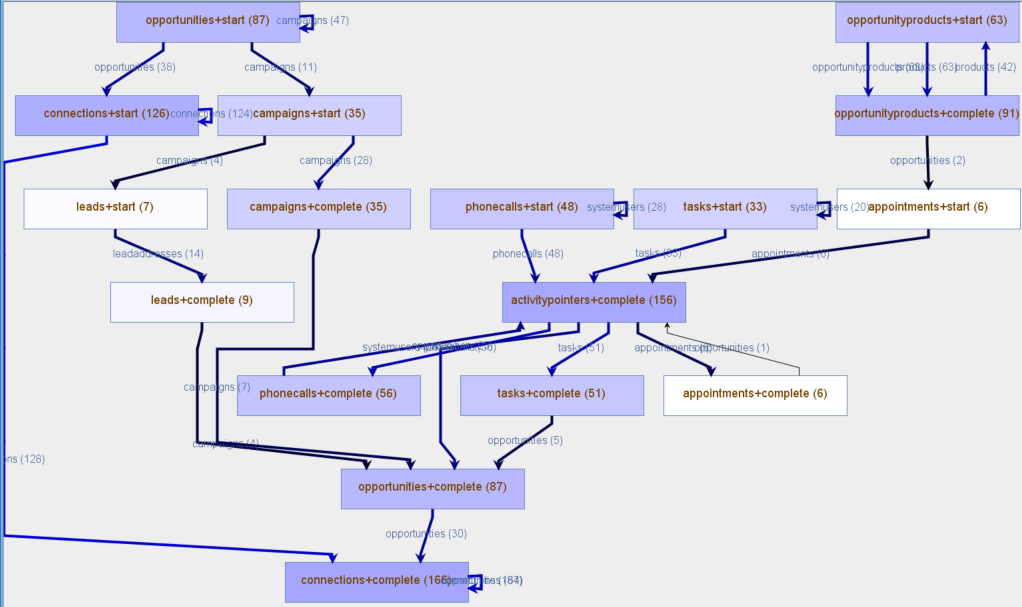}
\caption{Representation of a specific subset of activities in the A2A multigraph of the StarStar model extracted from a Dynamics CRM system as shown by the ProM plug-in.}
\label{fig:xocOpportunities}
\end{figure}
The previous construction means that for the same entry there are two events and one object in the XOC log. Moreover, in an XOC log events related to all the processes
belonging to the Dynamics CRM information system have been stored. The XOC log is stored inside the {\it tests} folder of the StarStar models package in order to provide a real use case of StarStar models.
The log could be imported using the StarStar model import plug-in and a visualization of the commercial opportunities processes and their relationships is provided.
The weight threshold is automatically set by the algorithm to $0.5$. For the process described by the log, a weight threshold of $0.4$ represent the model at a more appropriate level of complexity.
The most important processes of the commercial lifecycle are shown as a single connected component, describing the order in which activities have been performed:
\begin{itemize}
\item A commercial opportunity is started on the CRM system, associated to a lead (a potential customer that has been identified with enough elements to start a sales approach).
\item One or more of the following activities have been performed:
a set of products from the company catalog is chosen for the lead in order to propose a solution;
an ad-hoc marketing campaign, that could be targeted to companies of the sector or having the same issues/concerns, could be started;
a connection with other customers could be inserted in the system by sales people. This could be useful to better evaluate and qualify the lead.
\item Phone calls, appointments, or other types of commercial tasks (e.g. writing a message or involving the customer in a demo) can be set-up in the CRM.
\end{itemize}
A view of this process could be seen in Fig. \ref{fig:xocOpportunities}.
In Fig. \ref{fig:xocOpportunities}, information related to several entities (from the OData endpoint: {\it opportunity}, {\it opportunityproducts}, {\it campaigns}, {\it activitypointers}, {\it connections}) has been merged
in a single multigraph that takes into account several perspectives. Obtaining a classic event log describing all this information would have required to do a join of $5$ different database tables, while
obtaining the view was straightforward with StarStar models ProM plug-in. A problem of the A2A multigraph is that it does not capture concurrency / parallelism between activities; it is
required to extract a classic event log from the StarStar model and then do the analysis using a classic process discovery technique (like Inductive Miner). Applying the projection technique described in Section \ref{sec:projectionExpl}, choosing {\it opportunity} as perspective, with log window set equal to $2$ and connection weight threshold set to $0.05$,
and applying the Inductive Miner with noise threshold set to $0.4$, the Petri net displayed in Fig. \ref{fig:petriNet} is obtained.
The Petri net in Fig. \ref{fig:petriNet} provides further information about the order in which activities are done. In particular, it should be noted that:
\begin{itemize}
\item No activities related to campaigns are present in the Petri net. In general, campaigns can exist without opportunities,
and opportunities can exist without campaigns, so the connection threshold between them is very low.
\item The completion of commercial tasks may also happen after the event of closure of the opportunity. This is a problem in reporting: while tasks are
already performed, their documentation in the Dynamics CRM may happen sometime after the closure.
\end{itemize}
Another interesting process, that is contained in the Dynamics CRM but is not related to the core commercial activities, is the measurement of performance of the business.
In Dynamics CRM, there are {\it goals} (targets in terms of sales/contacts/campaigns diffusion) and {\it metrics} (how to effectively measure performance). Using the projection
function to obtain a classic event log (with log window set to $2$ and connection weight threshold set to $0.05$), and applying Inductive Miner with noise threshold
set to $0.2$, the Petri net displayed in Fig. \ref{fig:petriNet2} is obtained, where the setting of metrics usually happen before the setting of goals, and several goals could be defined using the same metrics.
The discovery of this process using classic queries on the Dynamics CRM database would have required the merge of at least two different entities. Even if this process
does not belong to the core commercial activities, StarStar models with no effort can present a complete view of all the processes contained in the Dynamics CRM.
\section{Conclusions}
\label{sec:conclusions}
This paper introduces StarStar models, providing a way to reduce ETL efforts on databases in order to enable process mining projects. StarStar models provide
a multigraph visualization of the relationships between activities happening in a database, and the possibility to drill down.
By selecting any case notion interactively we get a classic event log that can be analyzed using existing process mining techniques.
Each step in the construction of a StarStar model has linear complexity and can be done on graph databases (since all involved objects are graphs).
A plug-in has been implemented on the ProM framework that can import data extracted from databases (in XOC or SLEX format), build the StarStar model and provide a visualization
of the activities multigraph along with filtering and projection capabilities. While the activities multigraph does not provide a formal execution semantics, projected logs can be
used to extract Petri nets, so formal models with a clear execution semantic. Assessment done on the Dynamics CRM system gives evidence that StarStar models can handle
the complexity of current information systems, and provide an usable and time-effective ETL scenario.

%
%
%
%
\bibliographystyle{splncs04}

\end{document}